\documentclass[aps,showpacs]{revtex4}
\usepackage[dvips]{graphicx}
\usepackage{amsmath, amsthm, amssymb}
\usepackage[utf8]{inputenc}

\newcommand{\be}{\begin{equation}}
\newcommand{\ee}{\end{equation}}
\newcommand{\bea}{\begin{eqnarray}}
\newcommand{\eea}{\end{eqnarray}}
\newcommand{\bes}{\begin{subequations}}
\newcommand{\ees}{\end{subequations}}

\newcommand{\bb}{\bibitem}

\newcommand{\Ref}[1]{(\ref{#1})}
\def \a{\alpha}
\def \b{\beta}
\def \c{\gamma}

\def \L{{\cal L}}
\def \pr{\prime}

\def \arcsinh{\mathop{\rm arcsinh}\nolimits}

\begin{document}
\title{f(R)-Brane}
\author{V.I. Afonso, D. Bazeia, R. Menezes, and A.Yu. Petrov}
\affiliation{{Departamento de F\'{\i}sica, Universidade Federal da Para\'{\i}ba, 58051-970, Jo\~ao Pessoa PB, Brazil}}

\begin{abstract}
We investigate the existence of brane solutions in braneworld scenarios described by real scalar field in the presence of modified $f(R)$ gravity with a single extra dimension. In the case of constant curvature, we obtain first-order differential equations which solve the equations of motion and ease the search for explicit analytical solutions. Several examples of current interest are investigated to illustrate the results of the present work.
\end{abstract}

\maketitle
 
{\bf 1. Introduction.} The brane theory is being intensively investigated as a sound candidate for solving the hierarchy problem, and other fundamental problems
in high energy physics \cite{RS,RSa1,RSa2,RSa3,RSb}.
   
The brane concept which was first introduced in \cite{RS} is formulated in terms of a single infinite extra dimension according to which
the physical world appears as a four-dimensional spacetime embedded into an anti-de Sitter $(AdS)$ space of one higher spatial dimension.
This scenario starts with the five dimensional spacetime $AdS_5,$ with warped geometry described by a single function which is real and depends
only on the fifth or bulk coordinate, the extra dimension, which is an infinite spatial coordinate. The scenario is admissible under the presence of scalar fields which are also supposed to depend only on the extra dimension and work to stabilize the braneworld configuration \cite{RS,RSa1,RSa2,RSa3,RSb,csaki}. 
We name this the standard braneworld scenario, since it is governed by some Einstein-Hilbert action which includes gravity and scalar fields in a standard way. 
Similar ideas can also be found in \cite{DWC}.
 
The development of the brane concept was done almost simultaneously with other investigations, initiated with the observations of \cite{EO}, which has led to the modern concept of dark energy, a generic name used to identify the need of an extra but very important component of energy density to fill the spacetime and give rise to the present acceleration of the Universe -- detailed investigations can be found in the recent revisions \cite{R}.

We recognize as the standard cosmological model the $\Lambda$CDM model, which describes standard gravity with the scalar curvature $R,$ with the inclusion of the cosmological constant $\Lambda$ and the nonrelativistic dustlike matter. The presence of dark energy may be considered with diverse modifications of this standard scenario. The two main directions are {\it the quintessence way}, in which one considers the possibility of changing the standard scenario with the inclusion of scalar fields, and {\it the $f(R)$ way}, in which one modifies gravity, changing $R\to f(R),$ where $f(R)$ stands for a given function of the scalar curvature.

The recent interest on $f(R)$ gravity models \cite{fr} has increased very radically, and this has inspired us consider the issue of investigating the $f(R)$ modification of the braneworld scenario implemented in \cite{RS}. The subject of the present work is then a natural idea, which consists in the search of brane solutions with the $f(R)$ modification of gravity. Even knowing that this problem is hard to solve on general grounds, we believe that our recent advances on the first-order framework \cite{Ba} applied to study scalar fields coupled to gravity may help us to find explicit solutions. 

Other ideas of modifying the scenario of \cite{RS} have appeared in \cite{T,GB}, with the change of the scalar field dynamics to the tachyonic dynamics \cite{T}
or the inclusion of a Gauss-Bonnet term \cite{GB}. Although our proposal of including a $f(R)$ modification of gravity is different, it will certainly contribute to the subject.

{\bf 2. The framework.} The standard braneworld scenario is described by real scalar field $\phi$ interacting with gravity via the usual Einstein-Hilbert action, which has the general form
\be
S=\int d^4xdy \sqrt{g}\left(-\frac{1}{4}R+{\cal L}\right)
\ee
where we are using standard notation, and taking $4\pi G=1$. The scalar field $\phi$ is governed by standard dynamics, with Lagrange density given by
\be\label{model}
{\cal L}=\frac12 g_{ab}\partial^a\phi\,\partial^b\phi-V(\phi)
\ee
where $g_{ab}$ describes the five dimensional spacetime, with $a,b=0,1,2,3,4$ and $x^4\equiv y$ standing for the extra dimension.
The line element is given by
\be\label{metric}
ds^2_5=e^{2A}\,\eta_{\mu\nu}dx^\mu dx^\nu -dy^2
\ee
where $e^{2A}$ is the warp factor, and $A=A(y)$ a real function of the extra dimension which gives rise to the warped geometry and $\eta_{\mu\nu}=diag(+---)$ describes the four-dimensional flat spacetime, with $\mu,\nu=0,1,2,3$. The geometry of the five dimensional spacetime is then described by $A(y),$ and is driven by the extra coordinate $y$ alone. We name this the standard flat brane scenario.

In this work we maintain the metric \eqref{metric} and scalar field dynamics \eqref{model}, but we consider the $f(R)$ modification of gravity. This means that in the above action we change $R\to f(R)$ to get to the modified action
\be\label{ma}
S'=\int d^4xdy \sqrt{g}\left(-\frac{1}{4}f(R)+{\cal L}\right)
\ee 
which will be the action to govern the modified braneworld scenarios to be considered below. 

The five dimensional metric given above implies that the scalar of curvature depends only on the {\it fifth} coordinate $y,$ that is, $R=R(y).$ It can be written in terms of $A=A(y)$ in the form  
\be\label{curv}
R(y)= 8A^{\prime\prime}+ 20 A^{\prime}\,^2,
\ee 
where the prime denotes derivative with respect to $y$. 

We consider the modified action \eqref{ma} and suppose that the scalar field only depends on the extra dimension, that is, $\phi=\phi(y).$ Variation of the modified action with respect to the scalar field leads to the equation of motion 
\be
\phi^{\prime\prime}+4A^{\prime}\phi^{\prime}-\frac{dV(\phi)}{d\phi}=0 \label{eomphi}
\ee
which is the standard equation. However, the $f(R)$ modification changes the Einstein equations and we now get to
\be
R_{ab}f_{R}-\frac{1}{2}g_{ab}f(R)+(g_{ab}\nabla^2-\nabla_{a}\nabla_{b})f_R = 2\,T_{ab},\label{ee1}
\ee
where $f_R$ represents the R-derivative of $f,$ that is, $f_R={df(R)}/{dR}.$ We are also using 
\be 
T_{ab}=\frac{2}{\sqrt{g}}\frac{\delta(\sqrt{g}\,\L)}{\delta g^{ab}}
\ee
as the energy-momentum tensor corresponding to the ${\cal L}$ given by Eq.~\eqref{model}.

The Christoffel symbols associated to the metric \eqref{metric},  have non zero components 
$\Gamma^\a_{\b4}=A^{\prime}\delta^\a_\b$ and $\Gamma^4_{\b\c}= \eta_{\b\c}e^{2A}A^\pr$. 
As the $A$ field depends only on $y,$ the components of the Ricci tensor are $R_{\mu\nu}=\eta_{\mu\nu}e^{2A}(A^{\prime\prime}+4 A^{\prime2})$, $R_{44}=-4(A^{\prime\prime}+A^{\prime2}),$ and $R_{4\mu}=R_{\mu4}=0.$ Thus, the Ricci scalar takes the value given in Eq.~\eqref{curv}. 
Also, the components of the energy-momentum tensor are given by
$T_{\mu\nu}= \eta_{\mu\nu} e^{2A}(\frac12\phi^{\pr2}+V(\phi))$, 
$T_{44}=\frac12\phi^{\pr2}-V(\phi)$ and $T_{4\mu}=T_{\mu4}=0$.

We substitute the above results into \eqref{ee1} to obtain the two equations
\bes
\bea
\frac14(A^{\pr\pr}+4A^{\pr2})f_R -\frac18 f(R)-(\frac34 A^\pr f_R^\pr+\frac14 f_R^{\pr\pr})
&=&\frac14 \phi^{\pr2}+\frac12 V(\phi)\label{eea}\\
(A^{\pr\pr}+A^{\pr2})f_R-\frac18 f(R)-A^\pr f_R^\pr&=&-\frac14\phi^{\pr2}+\frac12 V(\phi)
\eea\ees
or better
\bes\label{ee2}
\bea
A^{\pr\pr}f_R-(\frac13A^\pr f_R^\pr -\frac13 f_R^{\pr\pr})&=&-\frac23\,\phi^{\pr2}\label{ee2a}\\
(A^{\pr\pr}+A^{\pr2})f_R-\frac18 f(R)-A^\pr f_R^\pr&=&-\frac14\phi^{\pr2}+\frac12 V(\phi)\label{ee2b}
\eea\ees
These are the Einstein's equations, written appropriately to the present study; if we take $f(R)=R$ and use \eqref{curv} they become
\be\label{aa}
A^{\pr\pr}=-\frac23\,\phi^{\pr2} \qquad \mbox{and} \qquad A^{\pr2}=\frac16\,\phi^{\pr2}-\frac13\,V(\phi)
\ee
which are just the equations we usually investigate in the standard scenario \cite{RSb}.

The above equations \eqref{eomphi} and \eqref{ee2} are the equations we have to deal with to find solutions in the modified braneworld scenario. In the standard scenario, we know that there are first-order differential equations which can be applied to solve the equations \eqref{eomphi} and \eqref{aa}. Thus, in the next section we consider some specific possibilities of writing first-order equations in the modified case, governed by \eqref{eomphi} and \eqref{ee2}. 

{\bf 3. The constant curvature case.} Let us first consider the constant curvature case. We have that $R^\pr=0$ and this reduces the equations \eqref{ee2}
to the form
\bes\bea
3f_R A^{\pr\pr}+2\phi^{\pr2}&=&0\label{eephi}\\
2 A^{\pr2}f_R-\frac14 f(R) -\frac56 \phi^{\pr2}&=&V(\phi) \label{eeV}
\eea\ees

In first place we describe the case of an empty space. We consider $\phi=\bar\phi=$ const, with $V(\bar\phi)=-1/L^2$, for $L$ being a lenght scale.
These equations are thus reduced to
\bes\bea
3f_R A^{\pr\pr}&=&0\label{eephi1}\\
2 A^{\pr2}f_R-\frac14 f(R)+\frac1{L^2} &=&0 \label{eeV1}
\eea\ees
If $f_R\neq 0$, we find that $A^{\pr\pr}=0;$ thus, the brane solution is $A(y)=-|y|/L;$ the singularity at the brane location is the same of the standard
case -- see, e.g., \cite{RSa2}. Thus, $R=20/L^2.$ As a result we get the following equation for the curvature
\bea
Rf_R-\frac{5}{2}f(R)+\frac{10}{L^2}=0.
\eea
This equation can be treated in two ways: (i) we suppose that $f(R)=R+\gamma R^n$ in order to arrive at the equation $({n}-{5}/{2})\gamma R^n-(3/2)R+{10}/{L^2}$=0, from which we can find $R;$ (ii) we can treate this equation as a differential equation for $f(R)$, whose solution is
\bea
f(R)=\frac{4}{L^2}+C R^{5/2}.
\eea

In a more general case, let us now consider $R$ constant in the presence of the scalar field. We thus implement a first-order framework to solve the system. It is inspired in the investigations \cite{Ba}, and is given by the suggestion that
\be\label{asatz}
A^\pr(y)= W(\phi(y)). 
\ee
This establish an explicit relationship between $A$ and $\phi$ via the function $W=W(\phi)$.
Then, using the equations above we can write 
\be\label{dphi}
\phi^\pr(y)=-\frac32 f_R\; W_\phi
\ee
This pair of first-order equations allows to obtain
\be
A^{\pr\pr}=-\frac32 f_R\; W_\phi^2
\ee
and the potential 
\be
V(\phi)=-\frac14 f(R)+2f_R\;W^2-\frac{15}{8} f_R^2\;W_\phi^2
\ee

In principle, $W$ is assumed to be an arbitrary function. However, the form of the metric (\ref{metric}) restricts it severely. In fact, using the ansatz (\ref{asatz}), the scalar curvature $R$ can be written as
\be\label{eq11}
R=8A^{\pr\pr}+20A^{\pr2}= -12f_R W_\phi^2+20 W^2
\ee
Therefore, $W$ has to obey the constraint
\be\label{eqW}
W_\phi^2-\frac{5}{3f_R}W^2+\frac{1}{12f_R}R=0
\ee
It shows that the form of $W$ depends explicitly on the specific scenario, that is, on the specific modification introduced by $f(R)$.

If we now use \eqref{dphi} and \eqref{eqW} we can write
\be\label{intphi}
\int\frac{d\phi}{W_\phi}=\int\frac{d\phi}{\sqrt{\beta_1 W^2-\beta_2}}=-\frac32 f_R y + c,
\ee
where we have set $\beta_1=5/(3f_R)$, $\beta_2=R/(12f_R)$ and $c$ is an integration constant. 
We see that for obtaining the field $\phi(y)$ we need to first calculate $W(\phi)$, and then integrate and invert Eq.~\eqref{intphi}, which may make the problem very hard to solve. 

To study Eq.~\eqref{eqW} it is convenient to write it as
\be 
W_\phi^2-\b_1W^2+\b_2=0,
\ee
which allows us identify the following characteristic cases:
\begin{enumerate}
\item Zero curvature ($R=0$). \begin{enumerate}
\item Case $f_R>0$ (or $\beta_1>0$ and $\beta_2=0$). In this case we have to solve 
\be\label{eqW0}
\frac{dW}{d\phi}=\pm\sqrt{\beta_1 W^2}.
\ee
This equation can be integrated to give
\be
W(\phi)=\exp(\pm\sqrt{\beta_1}(\phi-b_1)),
\ee
and the corresponding field solution reads
\be
\phi(y)= \mp\frac{1}{\sqrt{\beta_1}}\ln(\mp(5/2\,y-c\beta_1)) +c_1.
\ee
Here $b_1$ and $c_1$ are integration constants.

\item Case $f_R<0$ (or $\beta_1<0$ and $\beta_2=0$). This case presents no real solutions, as 
$\beta_1<0$ in equation \Ref{eqW0} appears inside a square root.
\end{enumerate}

To obtain the field $A(y)$ we can directly solve the equation $R=8A^{\prime\prime}+20 A^\pr\,^2=0$, which yields
\bea\label{r0}
e^{2A(y)}=a_1\left(y-y_0\right)^{\frac{4}{5}}
\eea
with $a_1$ and $y_0$ being integration constants.
\item Negative curvature ($R<0$).  \begin{enumerate}
\item $f_R>0$ (i.e. $\beta_1>0$ and $\beta_2<0$). We have
\be
\frac{dW}{d\phi}=\pm\sqrt{\beta_1 W^2+|\beta_2|}.
\ee
Integration of this equation yields
\be
W(\phi)=\pm\,2\,\beta\sinh(\sqrt{\beta_1}\phi+b_2),
\ee
and the field solution for this case is
\be
\phi(y)=\pm\frac{1}{\sqrt{\beta_1}}
\left(\arcsinh\left(\tan\left(\beta\,(5\,y-2\,c\,\beta_1)\right)\right)+c_2\right).
\ee
Here $b_2$ and $c_2$ are integration constants, and $\beta=\frac12\sqrt{|\beta_2|/|\beta_1|}$.

\item The case $f_R<0$ (or $\beta_1<0$ and $\beta_2>0$), is peculiar. Since $W(\phi)$ turns out to be purely imaginary, this case is not allowed.
\end{enumerate}

For the negative curvature case we can write the equation for $A(y)$ as $R=8A^{\prime\prime}+20 A^\pr\,^2=-|R|$. The solution yields
\be
e^{2A(y)}= \left(\frac{a_2^2}{|R|}\right)^{\frac{2}{5}} \left(\cos\left(\frac{1}{4}\sqrt{5\,|R|} (y-y_0)\right)\right)^{\frac{4}{5}}
\ee
where $a_2$ is an integration constant.
\item Positive curvature ($R>0$). \begin{enumerate}
\item $f_R>0$ (i.e. $\beta_1>0$ and $\beta_2>0$). We have 
\be
\frac{dW}{d\phi}=\pm\sqrt{\beta_1 W^2-\beta_2}.
\ee
This equation is easily integrated to give
\be
W(\phi)=\pm\,2\,\beta\cosh(\sqrt{\beta_1}\phi+b_3),
\ee
with $b_3$ being an integration constant. In this case, integrating and inverting equation \Ref{intphi} leads to inconsistent solutions for $\phi(y).$
\item $f_R<0$ (or $\beta_1<0$ and $\beta_2<0$). We have
\be
\frac{dW}{d\phi}=\pm\sqrt{|\beta_2|-|\beta_1|W^2},
\ee
which gives the solution
\be
W(\phi)=\pm\,2\,\beta\sin(\sqrt{|\beta_1|}\,\phi+b_4),
\ee
and the corresponding field solution
\bea
\phi(y)=\pm\frac{1}{\sqrt{|\beta_1|}}\left(\arcsin\left(\tanh(\beta\,(5\,y+2\,c\,|\beta_1|))\right)+c_4\right),
\eea
with $b_4$ and $c_4$ being integration constants.
\end{enumerate}

When $R$ is positive, we can find the warp factor by solving $R=8A^{\prime\prime}+20 A^\pr\,^2$. 
The solution for this equation looks like
\be
e^{2A(y)}=\left(\frac{a_4^2}{R}\right)^{\frac{2}{5}}\left(\cosh\left(\frac{1}{4}\sqrt{5\,R}\, (y-y_0)\right)\right)^{\frac{4}{5}},
\ee  
with $a_4$ being an integration constant.
\end{enumerate}

In the standard scenario, the presence of first-order equations is directly connected with supersymmetry and they nicely appear within the supergravity context \cite{RSa2,DWC}. In this sense, the first-order framework which we have just obtained suggests that the model engenders supersymmetric extension.

{\bf 4. Other procedures.} The first-order framework is very interesting, since it helps in the search of analytical solutions. However, its application in the form used in the previous section does not exhaust all possibilities. We then make further efforts below. In particular, we may propose specific ansatzes as a direct way to investigate the problem. In this case, it is better to consider the system of equations given by Eqs.~\eqref{eomphi} and \eqref{eea}, which we rename as
\bes\label{33}
\bea
\phi^{\pr\pr}+4A^{\pr}\phi^\pr-\frac{dV(\phi)}{d\phi}&=&0\label{em}
\\
2( A^{\pr\pr}+4 A^\pr\,^2)f_R - f(R)-(\,6A^\pr f_{R}^\pr +2f_{R}^{\prime\prime})&=&2\phi^\pr\,^2+4V(\phi)\label{effe}
\eea
\ees
We study this system for the specific choices which follow below.

{\bf 4.1 The case of constant curvature.}
If $R$ is constant, the equation \eqref{effe} is reduced to
\be
-f(R)+2\left(4 A^\pr\,^2+A^{\pr\pr} \right)f_R(R)=2 \phi^\pr\,^2+4V(\phi). \label{em2}
\ee

As the function $A(y)$ is determined by the condition on the scalar curvature $R$, 
it remains to find $\phi(y)$, $V(\phi)$ and $f(R)$ for all the three possibilities, $R<0,$ $R=0,$ and $R>0.$

Let us first consider the case $A(y)=-k|y|.$ For this $A$, we have  $R=20k^2$ and $A^{\prime\prime}=0$ for $y\neq 0$. Thus, after differentiating eq.~\eqref{em2} we can eliminate the potential term using (\ref{em}) to obtain an ODE for $\phi$
\be
\label{em3}
\phi^\pr (\phi^{\pr\pr}+2 A^\pr \phi^\pr)=0. \nonumber\\
\ee
Using that $A^\pr=k\, \mbox{sgn}(y)$ we obtain the solution
\be
\phi={\bar\phi_0} \,e^{-2k|y|}+C_\infty \;, 
\ee
with $\bar\phi_0$ and $C_\infty$ constants.
Therefore, from equation (\ref{em}), the consistent potential is
\be
V(\phi)=-2k^2(\phi-C_\infty)^2.
\ee

Because of the form of $A,$ the right-hand side of (\ref{em2}) turns out to be zero,
providing an equation for $f(R)$. Using that $A(y)=-k|y|$ and $R=20k^2$, this equation takes the form
\bea
f(R)-\frac{2}{5} R f_R=0,
\eea
which has the solution
\bea
f(R)=\a\, R^{5/2}.
\eea
with $\a=\,$const. Thus, we have found another solution for the constant positive curvature case.

We can also evaluate the zero curvature case. We study the case $f(R)=R+\gamma R^n$, so $f(R)=0$, but $f_R=1$ for $n>1$. The function $A(y)$ is given by (\ref{r0}) but, for simplicity, we choose $y_0=0$ using the translational invariance.
Therefore, we must solve the system
\bea\label{em1}
&&\phi^{\prime\prime}+4\frac{1}{\frac52y}\phi^{\prime}-\frac{dV(\phi)}{d\phi}=0;\\
&&2\phi^{\pr}\,^2+4V(\phi)=-\frac{3}{(\frac{5}{2}y)^2}.
\eea

A natural ansatz in this case is
\bea
\phi(y)=k\ln|y|,\quad\, V(\phi)=b\,e^{a\,\phi}.
\eea
Substituting this ansatz we find that the values of $k$, $a$ and $b$ are fixed as $k=3\pm\sqrt{3},$ $a=-2\,k^{-1},$ and $b=-3k/5$, while $\gamma$ remains a free parameter. 

There exists another solution for this zero curvature case (in which, however, it is difficult to express the potential $V$ as a function of $\phi$)
\bea
\phi(y)&=&\pm\sqrt{15}\left[(-1 + c_1 y^\frac{2}{5})^{\frac12}+\;
\mbox{arctan}\left((-1 + c_1 y^\frac{2}{5})^{-\frac12}\right)\right]+c_2,
\\
V(y)&=& \frac{9-15 \,c_1 y^\frac{2}{5}}{50y^2}.
\eea
\noindent Here $c_1$ and $c_2$ are integration constants.

{\bf 4.2. The case of non constant curvature.}
We can also try to solve the system \eqref{33} without restricting the curvature to a constant.
For the function $f(R)$ in the form $f(R)=R+\gamma R^n$ the system becomes
\bes\label{34}
\bea
\label{em4}
&&\phi^{\pr\pr}+4A^\pr \phi^\pr-\frac{dV(\phi)}{d\phi}=0;\\
&&-R-\gamma R^n-2\phi^\pr\,^2-4 V(\phi)+(8 A^\pr\,^2+ 2A^{\pr\pr})(1+n\gamma R^{n-1})-\nonumber
\\&&-6 n(n-1)\gamma A^\pr R^{n-2} R^\pr+2\left[n(n-1)(n-2)\gamma R^{n-3}
R^\pr\,^2+n(n-1)R^{n-2} R^{\pr\pr}\right]=0.
\eea\ees
Here we can substitute the curvature (\ref{curv}) together with the expressions
\bea
&&R^\pr=8A^{\pr\pr\pr}+40A^{\pr}A^{\pr\pr},\nonumber\\
&&R^{\pr\pr}=8A^{\pr\pr\pr\pr}+40 A^{\pr\pr}\,^2+40A^{\pr}A^{\pr\pr\pr}.
\eea
Then the system \eqref{34} reads
\bes\label{35}
\bea
\label{em5}
&&\phi^{\pr\pr}+4A^{\pr}\phi^{\pr}-\frac{dV(\phi)}{d\phi}=0;\\
&&-8A^{\pr\pr}-20 A^{\pr}\,^2-\gamma (8A^{\pr\pr}+20 A^\pr\,^2)^n-2 \phi^\pr\,^2 -4V(\phi)+\nonumber\\
&&+(8 A^\pr\,^2+2A^{\pr\pr})(1+n\gamma (8A^{\pr\pr}+20 A^{\pr}\,^2)^{n-1})
-\nonumber\\&&-6 A^\pr n(n-1)\gamma (8A^{\pr\pr}+20 A^\pr\,^2)^{n-2}[8A^{\pr\pr\pr}+40A^{\pr}A^{\pr\pr}]-\nonumber\\&-&2n(n-1)\gamma\left[(n-2)(8A^{\pr\pr}
+20 A^{\pr}\,^2)^{n-3}(8A^{\pr\pr\pr}+40A^{\pr}A^{\pr\pr})^2+\right.\nonumber\\
&&+\left.(8A^{\pr\pr}+20 A^\pr\,^2)^{n-2}(8A^{\pr\pr\pr\pr}+40 A^{\pr\pr}\,^2+40 A^{\pr}A^{\pr\pr\pr})\right]=0.\label{eem}
\eea\ees

This system can be solved using a variation of the first-order formalism used before. Let us suggest the following choices of ansatzes
\bea
\label{ans}
A^{\prime}=Ce^{m\phi},\quad\, \phi^{\prime}=De^{m\phi}. 
\eea
This is convenient since it allows to decouple the first equation (\ref{em5}) which is now treated as a differential equation for the potential. We note that it follows naturally from the above choices that $A=\lambda\phi$, with $\lambda={C}/{D}$ being a constant.
One can find $A^{\prime\prime}=CDme^{2m\phi}$, $A^{\prime\prime\prime}=2CD^2m^2e^{3m\phi}$, $A^{\prime\prime\prime\prime}=6CD^3m^3e^{4m\phi}$, $\phi^{\prime\prime}=D^2me^{2m\phi}$. Substituting these expressions into (\ref{em5}) leads to 
\bea
(D^2m+4CD)e^{2m\phi}-\frac{dV}{d\phi}=0,
\eea
which is easily integrated to give
\bea
\label{ep}
V(\phi)=\frac{1}{2m}(mD^2+4CD)e^{2m\phi},
\eea
which has an exponential behavior, of current interest to cosmology -- see e.g., \cite{R}.

We substitute the above choices (\ref{ans}) into the second Eq.~\eqref{eem}, and we use the exponential potential (\ref{ep}) to get to the following equation
\bea
&&e^{2m\phi}(-8CDm-20C^2)-\gamma(8CDm+20C^2)^ne^{2nm\phi}-\left(4D^2+\frac{8CD}{m}\right)e^{2m\phi}+\nonumber\\
&&+(8C^2+2CDm)[e^{2m\phi}+n\gamma(8CDm+20C^2)^{n-1}e^{2nm\phi}]-\nonumber\\
&&-6Ce^{2nm\phi}n(n-1)\gamma(8CDm+20C^2)^{n-2}(16CD^2m^2+40C^2Dm)-\nonumber\\
&&-2n(n-1)\gamma e^{2nm\phi}\big[(n-2)(8CDm+20C^2)^{n-3}(16CD^2m^2+40C^2Dm)^2+\nonumber\\
&&+(8CDm+20C^2)^{n-2}(48CD^3m^3+120C^2D^2m^2)\big]=0.
\eea
Comparing the terms associated with $e^{2m\phi}$ and $e^{2nm\phi}$ (recall that $n\neq1$) and extracting the overall factors $D^2$ and $D^{2n}$, we find two algebraic equations for $m$ and $\lambda={C}/{D}$
\bea
&&(m+2\lambda)(3\lambda m+2)=0;\nonumber\\
&&-(8\lambda m+20\lambda^2)^n+n(8\lambda^2+2\lambda m)(8\lambda m+20\lambda^2)^{n-1}-\nonumber\\
&&-6\lambda n(n-1)(8\lambda m+20\lambda^2)^{n-2}(16\lambda m^2+40\lambda^2m)-\nonumber\\
&&-2n(n-1)\big[(n-2)(8\lambda m+20\lambda^2)^{n-3}(16\lambda m^2+40\lambda^2m)^2+\nonumber\\
&&+(8\lambda m+20\lambda^2)^{n-2}(48\lambda m^3+120\lambda^2m^2)\big]=0.
\eea
We see that there are two possible solutions for $\lambda$ in terms of $m$: $\lambda=-2/{(3m)}$ and $-m/2.$ Below we find a way to fix $\lambda$ as a function of $m.$ The corresponding solutions for the fields $\phi$ and $A$ following the choices (\ref{ans}) are given by
\bea
\label{sols}
\phi(y)&=&-\frac{1}{m}\ln(C_0-mDy),\nonumber\\
A(y)&=&-\frac{\lambda}{m}\ln(C_0-mDy),
\eea
where $C_0$ is an integration constant. The curvature in this case is
\bea
R=8A^{\prime\prime}+20(A^{\prime})^2=\frac{8mCD+20C^2}{(C_0-mDy)^2}.
\eea
The singularity shown at $y_0={C_0}/{mD}$ identifies the location of the brane. Of course, the singularities that we have found in the present work are not specific of the $f(R)$ modification here studied -- similar singularities also appear in the standard scenario -- see, e.g., Ref.~\cite{G}.

We now use $f(R)=R+\gamma R^n$ in Eq.~\eqref{ee2a} to get
\bea
A^{\prime\prime}(1+n\gamma R^{n-1})-\frac{1}{3}\gamma n(n-1)R^{n-3}[A^{\prime}R^{\prime}R-(n-2)(R^{\prime})^2-R^{\prime\prime}]=-\frac{2}{3}\phi^{\prime2}.
\eea
This equation and the solutions \eqref{sols} lead to another equation, which can be written with two kind of terms, one proportional to $1/{(C_0-mDy)^2}$ and the other to ${1}/{(C_0-mDy)^{2n}}$:
\be
\frac{T_1}{(C_0-mDy)^{2}}+\frac{T_2}{(C_0-mDy)^{2n}}=0.
\ee 
Thus, since $n\neq1,$ from $T_1$ we obtain $\lambda=-2/(3m),$ and so the second possibility $\lambda=-m/2$ has to be discarded. We then use $\lambda=-2/(3m)$ in $T_2=0$ to get
\bea
m=\pm\sqrt{\frac{1}{7-2n}\left(\frac{1}{1-n}-\frac{2}{3}\right)},
\eea
which leads to the possible values of $m,$ for several $n\neq1.$ 

{\bf 5. Ending comments.} In this work we have extended the standard braneworld scenario to include the $f(R)$ modification of gravity. This modification makes the problem very hard to solve, but in the case of constant curvature we have been able to find a pair of first-order equations which solve the second-order equations of motion, and we have investigated several models. The presence of first-order equations eases the investigation, and we have included explicit solutions for all the three cases of positive, null or negative curvature. 

As we know, in the standard scenario the presence of first-order equations is directly related to supersymmetry. Thus, the first-order framework here obtained suggests that the $f(R)$ modification introduced in the present work may engender supersymmetric extension. Another issue concerns stability of the investigated scenarios, with the $f(R)$ modification here introduced, which may be studied within the lines of the third work in \cite{Ba}. These and other issues are presently under consideration, and we hope to report on them in the near future.

{\bf Acknowledgments.} The authors would like to thank CAPES, CLAF, CNPq and PRONEX-CNPq-FAPESQ for partial support. The work by AYuP is
supported by CNPq-FAPESQ DCR program, CNPq project No. 350400/2005-9.

\end{document}